# Visual Data Analysis with Task-Based Recommendations

Leixian Shen[1] · Enya Shen[1] · Zhiwei Tai[1] · Yihao Xu[1] · Jiaxiang Dong[1] · Jianmin Wang[1]

**Abstract**
General visualization recommendation systems typically make design decisions for the dataset automatically. However, most of them can only prune meaningless visualizations but fail to recommend targeted results. This paper contributes TaskVis, a task-oriented visualization recommendation system that allows users to select their tasks precisely on the interface. We first summarize a task base with 18 classical analytic tasks by a survey both in academia and industry. On this basis, we maintain a rule base, which extends empirical wisdom with our targeted modeling of the analytic tasks. Then, our rule-based approach enumerates all the candidate visualizations through answer set programming. After that, the generated charts can be ranked by four ranking schemes. Furthermore, we introduce a task-based combination recommendation strategy, leveraging a set of visualizations to give a brief view of the dataset collaboratively. Finally, we evaluate TaskVis through a series of use cases and a user study.

## 1 Introduction

Data visualization is a promising way to facilitate data exploration. However, transforming the raw data into well-designed visualizations requires users to be familiar with both the dataset and the principles of effective visual encoding. Although the visual data analysis demand has nourished numerous visualization tools, they usually present a steep learning curve. Users need to engage in a tedious and time-consuming process to explore different design decisions through trial and error. To address this issue, a remarkable series of Visualization Recommendation (VisRec) systems have been proposed both in academia [20, 25, 34, 39, 62, 64] and industry(e.g., Tableau and Power BI) to facilitate the visualization authoring process.

Over the past two decades, a surge of VisRec systems emerged as a powerful tool for visual analysis. The existing VisRec systems mainly consist of two categories [65]: rule-based approaches and learning-based approaches. Rule-based approaches build upon empirical visualization design knowledge. For instance, APT [37], Show me [38], DIVE [27], SeeDB [58], and Voyager [62] leverage predefined perceptual rules to automatically generate visualizations. These rule-based approaches are intuitive to understand and can guarantee good explainability of the recommendations. However, they suffer from high requirements for domain expertise. Learning-based approaches learn from plenty of well-designed visualization examples to train models for VisRec, such as DeepEye [34], Data2Vis [20], VizML [25], and Table2Charts [64]. Learning-based approaches avoid complicated hand-crafted rules, but they lack good scalability and may fail to achieve comparable performance as rule-based methods due to the limited training data.

Although these VisRec systems can generate meaningful visualizations, they pay relatively little attention to the diversity of user intents. Previous studies have documented that the effectiveness of a visualization largely depends on the user's analytic task [14, 43, 46]. For a dataset, different

✉ Leixian Shen
slx20@mails.tsinghua.edu.cn

Enya Shen
shenenya@tsinghua.edu.cn

Zhiwei Tai
tzw20@mails.tsinghua.edu.cn

Yihao Xu
yh-xu18@mails.tsinghua.edu.cn

Jiaxiang Dong
djx20@mails.tsinghua.edu.cn

Jianmin Wang
jimwang@tsinghua.edu.cn

[1] Tsinghua University, Beijing, China

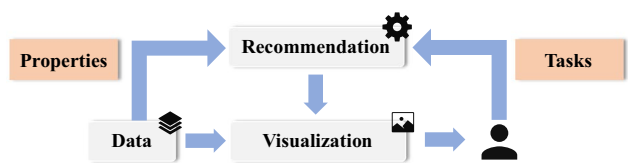

**Fig. 1** Recommendation pipeline. The recommendation engine accepts analytic tasks and data properties, and then automatically generates appropriate visualizations for users

people may hold different expectations of the recommendations according to their analysis intent. However, most of the existing VisRec systems lack detailed modeling of analytic tasks, so they can only prune meaningless results but fail to recommend targeted visualizations. As a result, given specific data columns, they will always produce the same charts for all the users. Even if some works have taken analytic tasks into account, the comprehensiveness and depth are still insufficient. For instance, Draco [39] only groups tasks into two categories: value tasks and summary tasks. NL4DV [40] only integrates five types of low-level analytic tasks into the VisRec system. Fortunately, there is a remarkable series of empirical wisdom on task modeling both in academia and industry. For example, academic work from Saket *et al*. [46] comprehensively evaluates the effectiveness of five commonly used chart types across different tasks. For industry, Financial Times Visual Vocabulary [11] provides great guidelines for designers and journalists to select the optimal symbology for data visualizations based on their tasks. However, all the design knowledge of analytic task are scattered across heterogeneous sources (e.g., papers, training sessions, projects, etc.) and are rarely implemented in practice to power visualization recommendation systems.

In response to the problem, we first summarize the empirical knowledge of the analytic task and then propose a task-oriented visualization recommendation system, TaskVis[1], for tabular data analysis. As shown in Fig. 1, the recommendation engine accepts analytic tasks and data properties and automatically generates appropriate visualizations for users. TaskVis differs from existing VisRec systems by modeling analytic tasks in detail. The contributions of this paper are summarized as follows:

- A task base of 18 common analytic tasks with their appropriate chart types by a survey both in academia and industry.
- A rule base, which extends empirical wisdom with our targeted modeling of analytic tasks.
- A task-oriented VisRec system, TaskVis, which allows users to select their analytic tasks precisely.

- A series of use cases and a user study to prove the effectiveness of TaskVis.

We note that this paper is an extended version of our earlier short paper in EuroVis 2021 [51].

## 2 Overview

This section gives an overview of TaskVis, including the preliminaries, user interface, and system architecture.

### 2.1 Preliminaries

This subsection provides the preliminaries of our approach, Vega-Lite and Answer Set Programming (ASP).

#### 2.1.1 Vega-Lite

Inspired by prior works [33, 39], we leverage Vega-Lite to specify visualizations. Vega-Lite [48] is a high-level language of interactive graphics. It provides a composition declarative JSON syntax for visualization displays. Vega-Lite supports various atomic components, including marks, channels, and transforms. Figure 2 shows an example of Vega-Lite specification, which is a bar chart, including the x-axis, y-axis, and color channel. In addition, sort, sum aggregation, and zero stack transformation are applied.

#### 2.1.2 Answer Set Programming

Answer set programming is a declarative programming paradigm based on logic programs and their answer sets. It provides a simple yet powerful modeling language to solve combinatorial problems. An ASP program is built on *atoms*, *literals*, and *rules*. *Atoms* are basic building blocks in ASP to represent elementary propositions. *Literals* are atoms A or their negation not A. A *rule* in ASP is in the form of A :- $L_0, L_1, ..., L_n$. where $L_i$ is a literal. The head ( A, left of :-) in the rule is true only if all the literals in the body ( $L_i$, right of :-) are true. For instance, road( cityA, cityC) :- road( cityA, cityB), road( cityB, cityC). states that only when city A and B are connected and city B and C are connected, city A and C can be connected. If a rule is headless, it means that satisfying all the literals in the body will lead to a contradiction. Illustrated with example, :- on( computer), off( computer). states that a computer cannot be both on and off. If a rule is bodyless, it asserts the fact that its head is true, for example, on( computer). states that the computer is on. Besides, the aggregate rule p $\{A_0, ..., A_n\}$ q. restrains that at least p and at most q atoms in the set are true. More modeling constructs

---

[1] The code is available at https://github.com/ShenLeixian/TaskVis.

**Fig. 2** Example of Vega-Lite specification. The figure shows a bar chart with an ordinal variable at x axis, a quantitative variable at y axis, and a nominal variable at color channel. In addition, sort, sum aggregation, and zero stack transformation are applied

```
{
  "$schema": "./vega-lite/v4.json",
  "data": {"url": "data/cars.json"},
  "mark": "bar",
  "encoding": {
    "x": {"type": "ordinal", "field": "Cylinders", "sort": "y"},
    "y": {
      "type": "quantitative",
      "aggregate": "sum",
      "field": "Horsepower",
      "stack": "zero" },
    "color": {"type": "nominal", "field": "Origin"}
  }
}
```

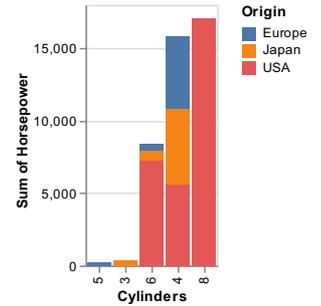

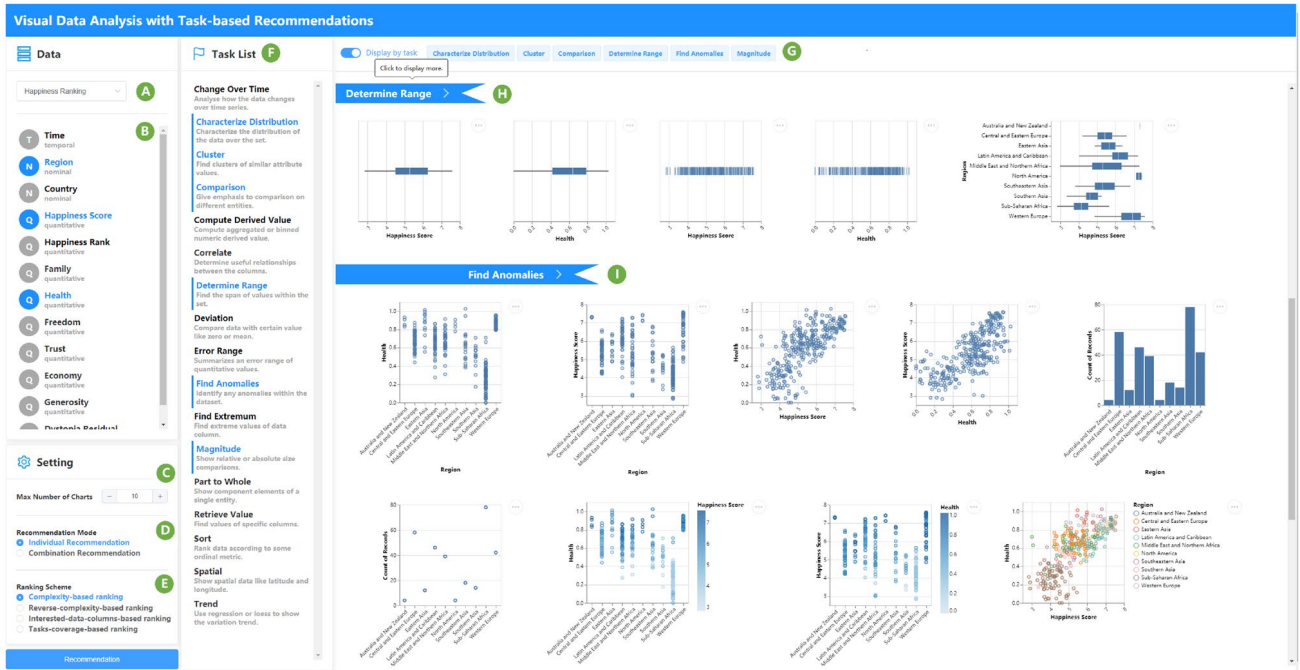

**Fig. 3** User interface. When uploading a dataset (**A**), the data field shows data columns with the corresponding data type (**B**). Users can customize system settings, including interested data columns (**B**), the max number of charts (**C**), recommendation mode (**D**), ranking scheme (**E**), and task list (**F**). Clicking the *Recommendation* button will generate visualizations on the right. Selected tasks are tabbed, along with a *Display by task* switch (**G**). If the switch is open, the charts will be displayed by tasks. (**H**) is the default thumbnail view with a row of charts. Clicking the task tab will show all the recommendations (**I**). If the switch is closed, all the charts will be deduplicated first and displayed together like in Fig. 6

of ASP can be found in its document[2]. Clingo [22] is an ASP system to solve logic programs. Given a combinatorial problem, a series of rules make up the ASP program, Clingo can be used as a solver to quickly enumerate all eligible combinations, with which the researchers can concentrate on the actual problem.

## 2.2 User Interface

The user interface of the system is shown in Fig. 3. After uploading the dataset (**A**), the data columns with the corresponding auto-detected data type (quantitative, nominal, ordinal, and temporal) are shown in the data field panel (**B**). Users can select data columns of interest by clicking multiple times on the panel. The setting panel allows users to customize recommendation configurations, including the max number of generated charts (**C**), two recommendation modes (**D**), and four ranking schemes (**E**). The task list panel with task description (**F**) is the most important interaction design in our system, which enables users to precisely select their analytic tasks. Different task choices and system configurations correspond to different recommended strategies. Clicking the *Recommendation* button will generate visualizations on the right. Selected tasks are tabbed for reference

---

[2] https://potassco.org/.

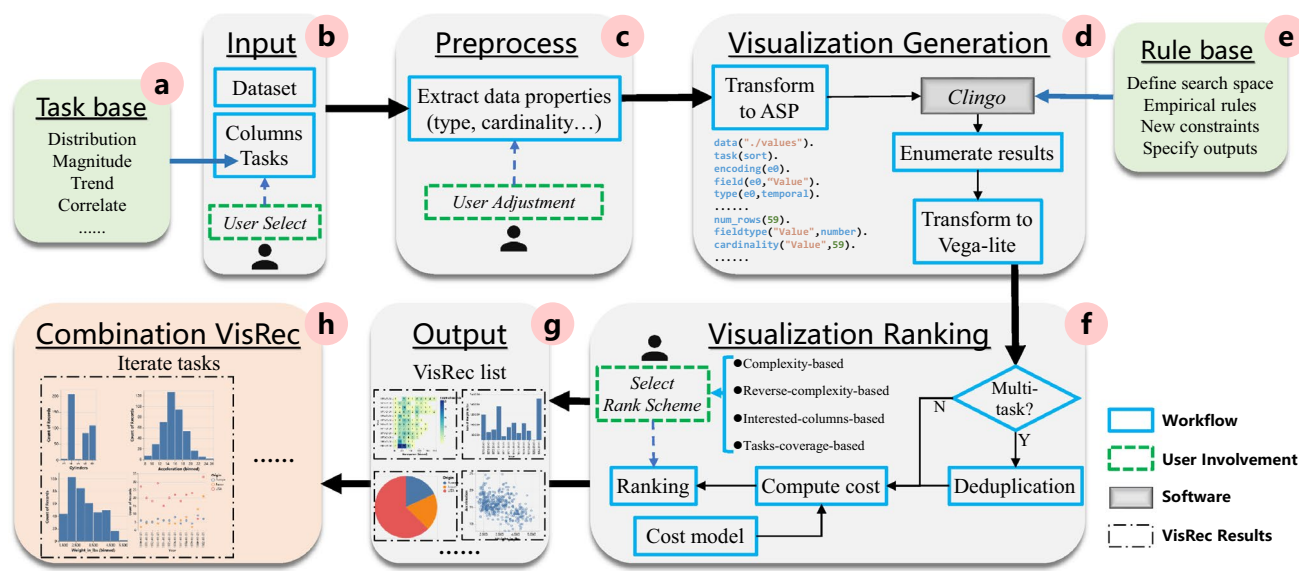

**Fig. 4** Architecture of TaskVis. TaskVis consists of six modules and two bases (for tasks **a** and rules **e** respectively). **b** Input: accepts the user's input. **c** Preprocess: extract data properties. **d** Visualization Generation: enumerate all qualified candidates. **f** Visualization Ranking: rank all visualizations according to selected scheme. **g** Output: present recommendation results to users. **h** Combination VisRec: make combination recommendations by iterating tasks

with a *Display by task* switch **(G)**. If the switch is open, the charts will be displayed by tasks, as shown in Fig. 3. **(H)** is the default thumbnail view with a row of charts. Clicking the task tab will show all the recommendations under a specific analytic task **(I)**. The chart can also be previewed by clicking on it like in Fig. 7. If the switch is closed, all the charts will be deduplicated first and displayed together as in Fig. 6. The analytic tasks to which the chart applies are marked at the top of the chart. The combination recommendation view is shown in Fig. 9.

## 2.3 System Architecture

The architecture of TaskVis is shown in Fig. 4. The task base **(a)** is the foundation of the system. A major difference between TaskVis and other VisRec systems is that TaskVis accepts analytic tasks as input in addition to the prerequisite dataset and optional data columns of interest **(b)**. So it can make targeted recommendations based on the task information. After inputting, the preprocessing module **(c)** will later extract data features (e.g., data type and cardinality) automatically. User intervention is also allowed in this phase as data types may be ambiguous in some scenarios. In the visualization generation module **(d)**, we adopt a rule-based approach, which can guarantee great explainability of the generated visualizations. The rule base **(e)** is the core of the generation module, which extends empirical visual design knowledge with our targeted modeling of analytic tasks. The rule base, along with the extracted data features and the user's input, is organized in ASP constraints. So we can further leverage Clingo, which is an ASP system to solve logic programs, to effectively enumerate all visualization candidates. All enumerated results in ASP constraints are transformed into Vega-Lite specification at the end of the visualization generation stage. In the visualization ranking module **(f)**, if previous modules have processed multi-tasks, the output VisRec list will be deduplicated. After that, each chart is assigned a *cost score* based on the cost model inspired by GraphScape [32]. On this basis, we design four novel ranking schemes for users to select, which can better satisfy users' diverse analysis requirements. Finally, a VisRec list **(g)** is presented to users, where each chart in the list is an independent recommendation. In addition, instead of just presenting individual charts, we propose a novel task-based combination recommendation strategy **(h)**, which leverages a set of charts to collaboratively describe the dataset based on task-oriented design principles.

## 3 TaskVis

In this section, we first introduce the modeling of analytic tasks. Then, we discuss the visualization generation and visualization ranking. Finally, we introduce the combination recommendation strategy.

**Table 1** Task base. *Mark* column lists appropriate marks, where the rank has priority, (*) indicates the combination of marks, e.g. rect(text) means a text layer is superimposed on rect chart

| Task | Mark | Description | Reference |
| --- | --- | --- | --- |
| Change Over Time | line/area | Analyse how the data changes over time series | [3, 6, 9–11, 24] |
| Characterize Distribution | bar/point | Characterize the distribution of the data over the set | [2–4, 6, 11, 13, 18, 40, 46, 47, 54, 60] |
| Cluster | bar/point | Find clusters of similar attribute values | [3, 13, 46, 52, 54] |
| Comparison | line/point/bar | Give emphasis to comparison on different entities | [2, 4, 6, 9, 10, 26, 27, 31, 47, 60] |
| Compute Derived Value | rect(text)/arc/bar | Compute aggregated or binned numeric derived value | [13, 27, 46, 54, 60] |
| Correlate | bar/line | Determine useful relationships between the columns | [2–4, 6, 9, 11, 13, 18, 24, 27, 40, 46, 47, 52, 54, 60] |
| Determine Range | tick/boxplot | Find the span of values within the set | [6, 13, 46, 54] |
| Deviation | bar(rule)/point(rule) | Compare data with certain value like zero or mean | [11] |
| Error Range | errorband/errorbar | Summarizes an error range of quantitative values | [15] |
| Filter | rect/bar/arc | Find data cases satisfying the given constrains | [13, 40, 46, 54, 60] |
| Find Anomalies | bar/point | Identify any anomalies within the dataset | [13, 18, 24, 26, 46, 47, 52, 54, 60] |
| Find Extremum | bar/point | Find extreme values of data column | [13, 26, 31, 46, 54, 60] |
| Magnitude | arc/bar | Show relative or absolute size comparisons | [6, 11, 53] |
| Part to Whole | arc | Show component elements of a single entity | [2, 4, 6, 9, 11, 53, 60] |
| Retrieve Value | rect(text) | Find values of specific columns | [3, 10, 13, 26, 31, 40, 46, 54, 60] |
| Sort | bar | Rank data according to some ordinal metric | [11, 13, 24, 46, 54, 60] |
| Spatial | geoshape/circle(text) | Show spatial data like latitude and longitude | [3, 6, 9–11] |
| Trend | point(line) | Use regression or loess to show the variation trend | [2, 27, 59, 60] |

## 3.1 Analytic Tasks

While many task taxonomies have been constructed, most of them only apply to specific scenarios. Besides, based on existing task taxonomies, some new tasks introduced by incremental works for visual data analysis are scattered across heterogeneous sources. To summarize these tasks, we maintain a task base from three aspects:

- *Empirical academic studies* A large body of early works pay attention to the effectiveness of visualization types for a selected task, either special for a particular chart type (e.g., scatter [47, 59], bar [16, 21], circle [16, 21], error bar [15], arc [53], area [53], and line [59]) or comprehensive evaluation [13, 18, 24, 31, 46, 60].
- *Empirical summaries in the industry* Various fields such as newspaper, finance, geography, and engineering have a strong requirement for data visualization, rich practice experience has also been accumulated in the form of guides, training sessions, projects, libraries, etc. [2–4, 6, 9–11].
- *Customized tasks* Apart from commonly used tasks, task base also contains some others that we consider meaningful during our visualization practice, such as *error range*.

Based on the studies above, similar to the methodology to identify scatterplot tasks in [47], we first formulate the seeds for the analytic task list, covering both low-level and high-level tasks. To abstract these analytic tasks, we then invite several data visualization researchers with more than five years of experience to perform a card sort and merge similar tasks. Finally, we summarize 18 classical analytic tasks and their appropriate chart types as shown in Table 1. What needs to explain is that the *filter* task requires users to specify some constraints of data additionally. So the implementation of this task is placed in the data preprocessing stage, allowing users to filter data according to their own needs. For the other 17 tasks, users can choose anyone to perform data analysis at will. We also extract appropriate chart types for each task and retained the priority. For instance, the most appropriate chart type for the *comparison* task is the line chart, followed by the scatter chart, and finally the bar chart. (*) in the table indicates the combination of marks. For example, *rect(text)* represents that a text layer is superimposed on the rect chart.

## 3.2 Visualization Generation

Generally, VisRec systems need to enumerate all possible visualizations first and then make recommendations [42].

### 3.2.1 Visualization Components

Vega-Lite includes various atomic components. We support a part of them as follows:

*Mark* To better assist data analysts, we enrich the chart types of recommendations as much as possible. To this end, we merge and filter all the marks support in Vega-Lite and finally select 14 types, arc, area, bar, box plot, circle, error band, error bar, geoshape, line, point, rect, rule, text, and tick. Besides, TaskVis also supports layered charts (e.g., line + point) in some tasks. Compared with existing VisRec systems, TaskVis can recommend richer charts.

*Channel:* In visual analysis, the rational application of visualization channels is one of the pivotal factors for designing aesthetic visualizations. Visualization channels include location, color, size, shape, slope, texture, and animation. There is a consensus that the first four can generally meet the demands of visualization design. Hence for simplicity, encoding channels in TaskVis only refer to the axis (x and y) and legend (color, size, and shape). It is worth mentioning that at most two legends are involved in a chart.

*Transform* Data transformation enables new value generation by various operations. For example, we can use *bin* to bucket quantitative or temporal data, discretizing continuous variables. *Regression* and *loess* can be applied to produce trend lines. All the transform operations of Vega-Lite included in TaskVis are bin, aggregate, sort, stack, regression, loess, and scale.

### 3.2.2 Rule Base

Various visualization components form a huge search space, among which there are numerous invalid combinations and meaningless visualizations that show limited information. Fortunately, there are numerous rules either from users or traditional wisdom to prune "bad" visualizations. However, if all these rules are implemented in code with branch structure, it will be complicated to manage. Inspired by Draco [39], we leverage answer set programming to construct design knowledge in a unified and extensible manner and finally build a rule base, which consists of three sources: empirical wisdom, customized rules for task, and user partial specification.

*Empirical wisdom* Draco [39] maintains a knowledge base in ASP depending on prior efforts such as APT [37] and CompassQL [61]. ASP program of the knowledge base includes visualization attributes declaration, preference over design space, and output specification. Visualization attributes declaration specifies the domains of attributes, for example, type ( quantitative; ordinal; nominal; temporal). defines the four high-level data types and channel ( x; y; color; size; shape). defines the five supported single encoding channels. Preference rules over design space can effectively prune the search space, for example, :- channel( E, shape), not type( E, nominal). ensures that the model will not consider applying the shape channel on non-nominal variables. Specification of outputs like # show channel/ 2 defines the output items and the number of its parameters. With the aforementioned modeling process, Draco can be leveraged to build increasingly sophisticated visualization systems with less effort. In addition to Draco, we also integrate appropriate rules from other sources as listed in Table 1.

*Customized rules for tasks* Draco [39] only includes two abstract tasks (value and summary), which can hardly capture the user's activities during visual analysis. In response, TaskVis extends Draco with additional detailed modeling of analytic tasks based on the task base in Sect. 3.1. TaskVis introduces an additional primitive task(). As shown in Table 1, we summarize appropriate chart types for each task. It can be formulated as rules to effectively narrow the search space, for example, :- task( sort), not mark( bar). is an integrity constraint stating that it cannot be the case that mark type applied in *sort* task is not bar. There is another type of rule that restricts operations to several specific tasks. For example, aggregation is not allowed in the *determine range*, *trend*, *deviation*, and *error range* task, as they should display the raw data. Besides, based on the characteristics of each task, we also formulate separate rules independently. For instance, :- channel( E, x), not type( E, temporal), task(change_over_ time). states that the variable on the x-axis in the *change over time* task must be temporal. We develop these rules through iterative informal feedback from researchers and students with varying-level experience with visualizations based on a prototype of TaskVis. In each round, we pick out meaningless charts and further improve our rule base.

*User partial specification* TaskVis supports the full-automatic recommendation strategy, users only need to input the dataset and the system can automatically generate appropriate visualizations. However, to make user input more flexible, we allow inputting partial specifications which include interested data columns and their analytic tasks. It helps to restrict the search space within well-formed specifications of the user's interests. These rules are generated based on interactions between users and the visualization interface, as shown in Fig. 3(B, F).

Recall Fig. 4d, after transforming to ASP programs, the user's input along with the rule base is sent into Clingo. Differ from Draco [39], which eliminates meaningless visualizations using hard constraints and finds the most preferred charts according to soft constraints. TaskVis first enumerates all candidate visualizations in ASP by hard constraints and then converts to Vega-Lite specification for the visualization ranking module to handle. Each selected task corresponds to an independent VisRec list.

### 3.3 Visualization Ranking

TaskVis allows users to input multiple tasks, and some tasks have overlapping recommendations. This means that one chart may apply to multiple tasks. Therefore, if the visualization generation module handles multi-tasks, the VisRec list can be deduplicated first. After that, we derive a cost model inspired by GraphScape [32] to quantify the visualizations and design four ranking schemes.

#### 3.3.1 Cost Model

As TaskVis is powered by Vega-Lite, a single visualization can be split into independent components, as shown in Fig. 2. GraphScape [32] introduces a graph structure for modeling relationships among Vega-Lite-based charts. Graph nodes represent visualizations and edges represent edits that transform one visualization into another. GraphScape leverages linear programs to derive transition costs via a partial ordering of visualization components. Illustrated with an example, in the transition category of visual encodings, following the perceptual kernels method [19], metric-space embeddings are performed on visual encoding type comparisons. The results find the following transition cost order: x, y >row, column >color >size >shape >text, which assumes that changes to spatial encodings require more interpretation effort.

Inspired by GraphScape [32], we also generate a cost model and define a *cost score* to measure the complexity of different components. For a chart, the *cost score* can be obtained by adding the corresponding cost values of all components. A lower *cost score* means the visualization is more concise. In detail, we extend GraphScape with the supplement and subdivision of visualization components. For mark type, we refer to Table 1 to generate cost with the priority of each task. Among channels, we replenish some components that are special for arc and geoshape charts. As to data transformation, we not only add operations like stack, regression, and loess, but also subdivide aggregate operations into AGGREGATE_SUM, AGGREGATE_ COUNT, and AGGREGATE_MEAN. The corresponding generated cost is shown in Fig. 5 (left). Based on the cost model, each visualization can be assigned a *cost score* with good explainability. For example, in Fig. 2, the chart is generated under the *sort* task. Its component set and corresponding cost are shown in Fig. 5 (right), *cost score* can be obtained by summing up the cost of all components.

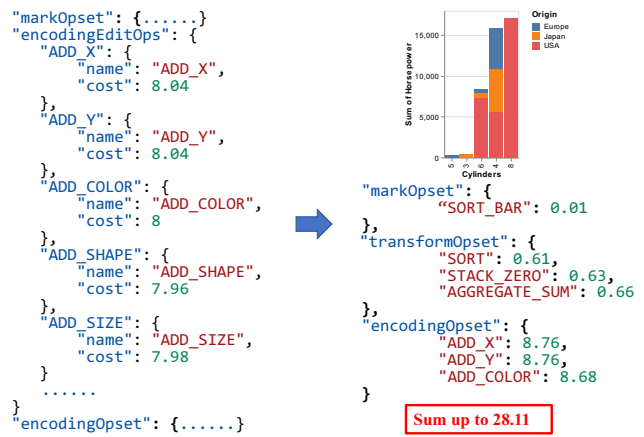

**Fig. 5** Example of cost model. *cost score* of the visualization in Fig. 2 can be obtained by summing up cost of all components

#### 3.3.2 Ranking Schemes

After computing the *cost score* of each chart, four ranking schemes are designed as follows:

(I) *Complexity-based ranking* This scheme ranks visualizations according to the chart's *cost score* from low to high. When people make visualization observations, they tend to view charts as a whole and hope to understand the information inside charts in a short time. According to our cost model, charts with lower *cost scores* are easier to understand, which is in line with people's habit of observing charts. This scheme is appropriate for most analytic tasks, such as *change over time*, *characterize distribution*, and *cluster*.

(II) *Reverse-complexity-based ranking* Overall, charts with higher *cost scores* come first in this scheme. However, instead of reversing the visualizations in the complexity-based ranking scheme, we first adopt the unsupervised algorithm DBSCAN to cluster all the results into several categories, where visualizations in one category are similar to each other (e.g., just exchange the coordinate axis or replace an aggregation function). Then the visualizations are ranked by category in reverse order while maintaining the partial rankings within the category. The scheme is appropriate for tasks like *sort* and *determine range*. The rationality is that charts of these tasks are relatively simple, charts with high *cost scores* can present more information within a range that humans can easily perceive.

(III) *Interested-data-columns-based ranking* This scheme is mainly related to the data columns of interest inputted by the user. Users usually expect a chart to display as many data columns of interest as possible. So if a chart fails to show all the user's interested columns, it will be assigned a penalty. The new *cost score* is computed by dividing (*N1/N2*), where N1 is the number of data columns covered in a single chart and N2 is the number of the user's interested data columns. Visualizations are finally sorted by the new generated *cost score* from low to high. The scheme fits some comprehensive tasks like *magnitude* and *find anomalies*.

(IV) *Tasks-coverage-based ranking* The above three ranking schemes are more suitable for single-task situations. For multiple tasks, we believe that if a chart is an overlap of multiple tasks' VisRec list, it must have universal meaning and should be ranked at the forefront. We define the *task coverage number* of a chart as the number of tasks (within the user's input) it can satisfy. Before ranking by this scheme, we merge and deduplicate all tasks' VisRec list first, and identical visualizations average the *cost score* value. On this basis, all charts are first ranked by *task coverage number*, where those with the same *task coverage number* are partially ranked by the *cost score*. Noting that this ranking scheme is valid only when TaskVis accepts multiple tasks. Otherwise, it will degenerate into the complexity-based ranking scheme.

Back to Fig. 4f, users are allowed to select the four ranking schemes freely to deal with the generated visualizations. We also recommend the appropriate ranking scheme for each task as default in the system based on the materials mentioned in Sect. 3.1. In the output VisRec list (g), each chart is an independent result.

---

**Algorithm 1** Combination Recommendation

**Require:** *Result*: Visualization class; *Result.Fields*: Columns involved in a chart; $Cols_{interest}$: Columns of interest that users input; *Recs* : *list*[*Result*]: Visualization list after deduplication.
**Ensure:** A set of visualizations *Ans* : *list*[*Result*];
1: sort Recs by Tasks-coverage-based ranking scheme;
2: set $Ans = Cols_{covered} = [\,]$;
3: **while** $Cols_{covered} \subset Cols_{interest}$ and *Recs* != *NULL* **do**
4:    append *Recs*[0] to *Ans*;
5:    *chosen fields* = *Recs*[0]*.Fields*;
6:    $Cols_{covered} = Cols_{covered} \cup chosen fields$;
7:    *enums* = all data column combinations in $Cols_{covered}$;
8:    **for** *vis* in *Recs* **do**
9:       **if** *vis.Fields* in *enums* **then**
10:          delete *vis* in *Recs*;
11:       **end if**
12:    **end for**
13: **end while**
14: **return** *Ans*;

---

## 3.4 Combination Recommendation

In this section, instead of recommending individual charts, we propose a task-based combination recommendation strategy, focusing on generating a set of visualizations to give a brief view of the whole dataset, as shown in Fig. 4h. This is the gospel of users who are not familiar with the data and it is an effective way for the system cold start. To this end, we extend the tasks-coverage-based ranking scheme in the following two situations:

*Without Task selection* If users select no tasks, the generation of candidate visualizations relies on iterating all tasks in the task base. Combination recommendations are further formulated according to the following principles: (1) The set of charts must cover all the data columns of interest to users. (2) Larger *task coverage number* is always chosen first. We describe the combination recommendation algorithm in Algorithm 1. The visualization candidates (*Recs*) are first ranked by the task-coverage-based ranking scheme. The algorithm always selects *Recs[0]*, which covers most analytic tasks. After that, visualization candidates with data column combinations that have been covered in the selection will be filtered. The algorithm iterates the above steps until all data columns of interest to users are covered. The time complexity of our algorithm is $O(n^2)$. Benefiting from the rule base effectively eliminating meaningless charts, the algorithm usually deals with a small set of visualization candidates and can meet the real-time analysis requirement.

*With Task selection* If users have selected tasks of interest, iterating the inputted tasks can generate candidate visualizations. The first principle mentioned above will be relaxed here because some data columns are only applicable for specific tasks. For example, the latitude and longitude variable can only appear in the *spatial* task. If only a single task is accepted, TaskVis can also recommend a set of targeted charts to describe the task.

## 4 Evaluation

This section will discuss the evaluation of TaskVis through a series of use cases and a user study.

### 4.1 Use Cases

This section will illustrate the recommendation capability of TaskVis with use cases. The individual recommendation will be exhibited first, and the combination recommendation follows.

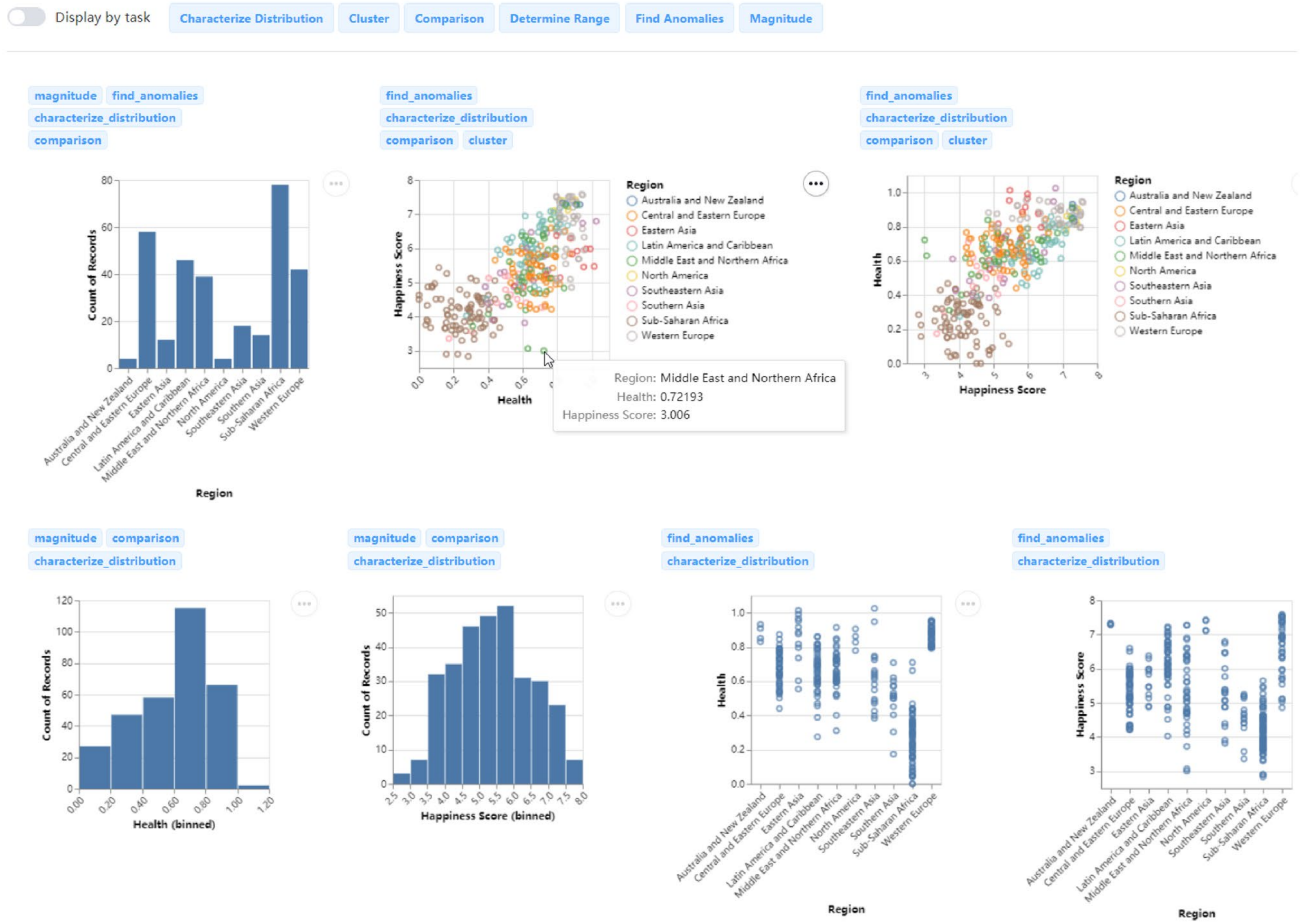

**Fig. 6** VisRec results of the Happiness Ranking dataset with the tasks-coverage-based ranking scheme. The *Display by task* switch is closed

### 4.1.1 Individual Recommendation

Imagine that we are exploring the Happiness Ranking dataset [12], after choosing the *Region*, *Happiness Score*, and *Health* columns, selecting six tasks, and closing the *Display by task* switch, all the charts are deduplicated and displayed together. As shown in Fig. 6, the tasks-coverage-based ranking scheme is applied, charts ranked in the forefront are appropriate for more tasks, and the names are placed above the chart. The charts also support simple interactions like hovering to see data details.

For the COVID-19 dataset of the U.S. [5], which contains the number of daily confirmed and dead cases, we may be interested in the geographical distribution of deaths. The recommendations for the *spatial* task are shown in Fig. 7. One chart is clicked for preview, it applies a color channel on the circle mark, and the concrete numbers are labeled. Geoshape charts leverage heat maps to reveal the distribution of deaths in each state. For this dataset, we may also be keen on "how the number of daily confirmed cases has changed since the epidemic?". After selecting the *change over time* task and data columns of interest, the recommendation results are shown in Fig. 8. We can intuitively observe the change in the total number of daily confirmed (national and four major regions distributed) by line and area charts, and the area charts also show the stack of data.

### 4.1.2 Combination Recommendation

Imagine that we are analyzing the Hollywood Stories dataset [7] and are not familiar with the data. A general view of the dataset can be helpful for our subsequent in-depth exploration. As shown in Fig. 9, without selecting tasks and data columns, the set of charts covers all the data columns and satisfies the most tasks. As the system displays thumbnails by default, high-cardinality variables (like *name*) will make the chart crowded, but clicking on the chart like in Fig. 7 can view the original clear chart. Although the use cases present promising results, our target is not to show that combination recommendation results are always superior. We argue that it offers an option to obtain a general view of the dataset.

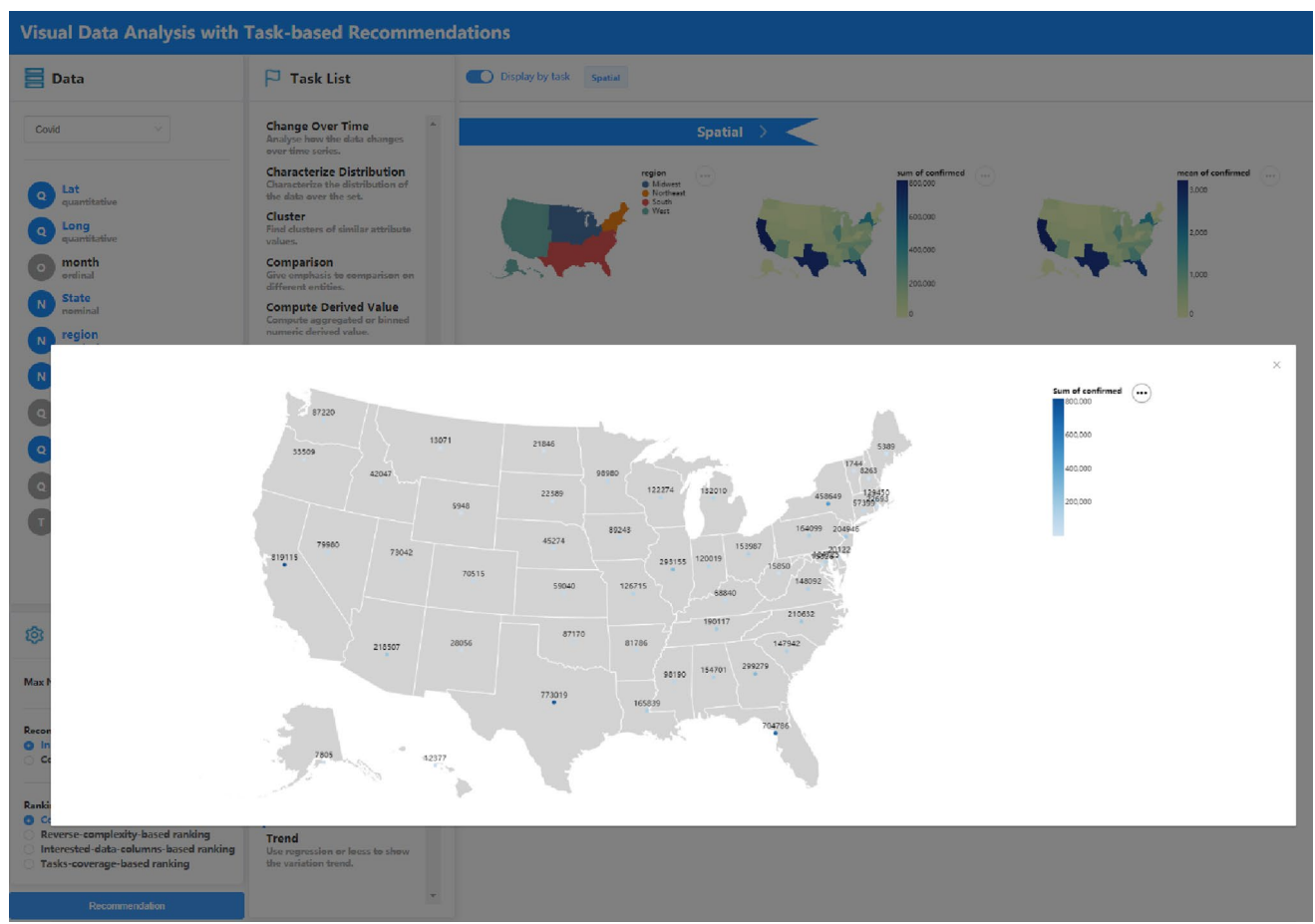

**Fig. 7** VisRec examples of the COVID-19 dataset in the *spatial* task with multiple columns. One chart is previewed by clicking on it

### 4.2 User Study

The overall goal of the study was to investigate TaskVis's ability to recommend targeted charts to satisfy the user's analytic tasks.

#### 4.2.1 Participants

We invited 22 persons (6 female and 16 male) to participate in the study. The participants include 1 undergraduate, 13 master candidates, 6 Ph.D. candidates, 1 researcher, and 1 data analyst. Among them, 8 participants perform data analysis every day, 9 monthly, and 5 less than once a month. All of them have the experience of using Microsoft Excel and programming with python to analyze and present their data.

#### 4.2.2 Study Protocol

The study can be roughly classified into three parts: pre-designed experiment, free exploration, and post-study interviews. Before the study, we presented a short examination of color blindness and basic visualization knowledge. In the pre-designed experiment, we first made an explanation of the process in detail and gave a brief description of each task. Then, given a recommended chart, participants were asked to answer a 5-point Likert scale question (-2 = strongly disagree, -1, 0, 1, 2 = strongly agree), "The chart of the {1} dataset can satisfy the {2} task.", where {1} is the dataset name and {2} is the task name. Metadata about the chart (e.g., task description and data columns) is also presented to help participants make their decisions. In total, there were 287 visualizations for participants to cope with. After the experiment, we made a tutorial to introduce the system interface and show some examples. Participants were later asked to freely explore TaskVis through a web browser on their computer, especially to perform tasks related to their actual day-to-day work. The user study lasted about one hour. Participants did not receive any intervention during the whole process. Finally, we conducted post-study interviews to get participants' feedback, including a questionnaire guided by the system usability scale to gauge the usability of TaskVis [8].

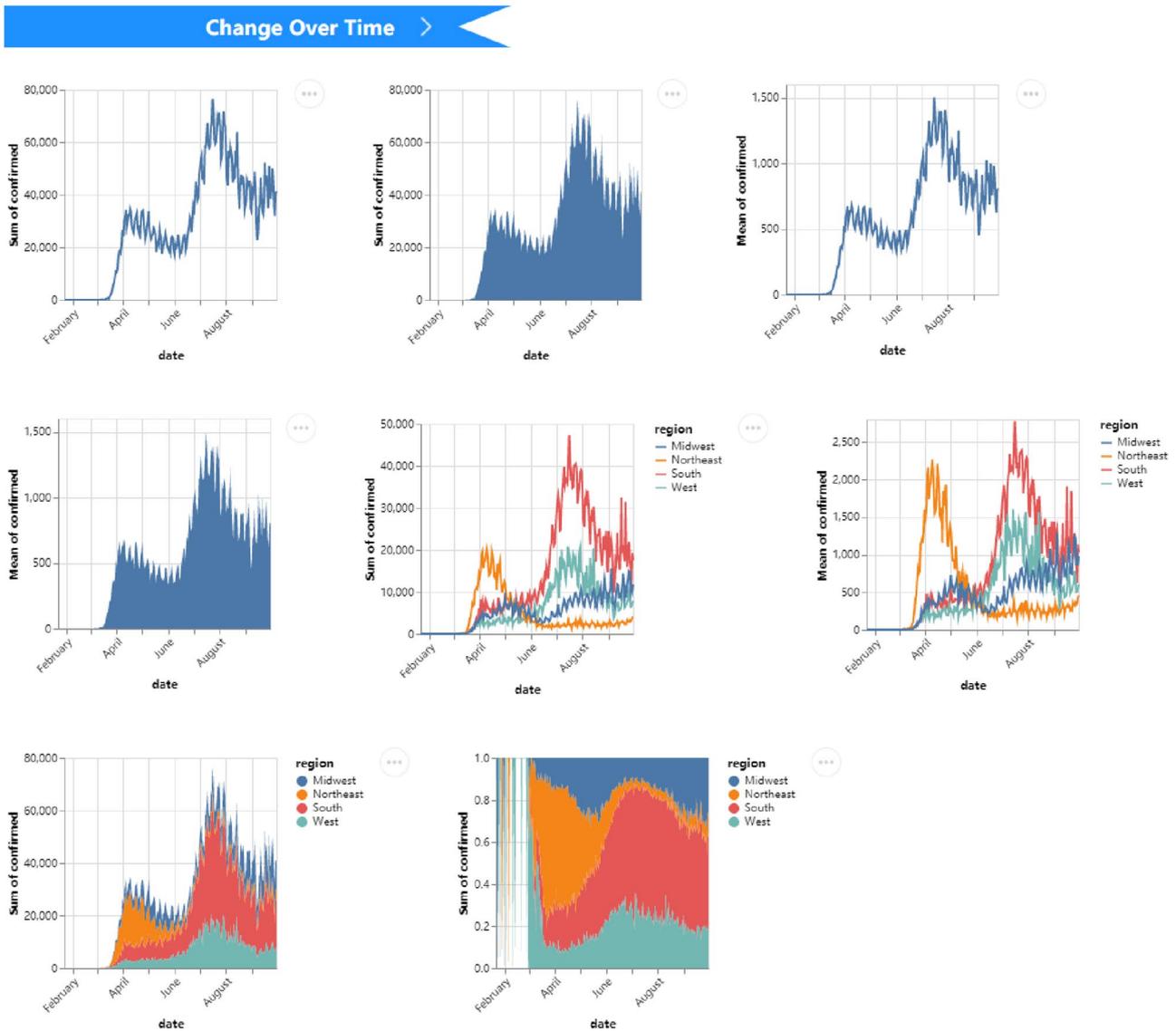

**Fig. 8** VisRec examples of COVID-19 dataset in the *change over time* task with the *region*, *confirmed*, and *date* columns

### 4.2.3 Datasets

We provided participants with two datasets in the pre-designed experiment. One is the COVID-19 data of the United States [5], including the daily numbers of confirmed cases and deaths in each state. The dataset has 12903 records and consists of 5 quantitative, 3 nominal, and 1 temporal columns. The other is sales data of 406 different cars [1], which is composed of 5 quantitative, 2 nominal, 1 ordinal, and 1 temporal columns. For free exploration, we offered more datasets covering areas of weather, sport, movie, etc.

### 4.2.4 Analysis of Result

All participants finished the pre-designed experiment and explored the system with several datasets. We finally collected 6249 valid records from 22 participants in the pre-designed experiment. The ratings with a 5-point Likert scale are shown in Fig. 10. The average score is 1.05. In general, TaskVis can recommend targeted charts to satisfy the user's analytic task. We further analyzed the charts that were scored -2 (strongly disagree) or -1 (disagree) more than three times and summarized the reasons as follows: (1) The overlap between marks makes charts difficult to distinguish, especially when a large amount of data is presented; (2) Outliers affect the users' understanding of the chart. For example, in the COVID-19 dataset, the number of deaths in one day exceeds 5000, which is much higher than the average. This may lead to some charts presented in extreme ways; (3) The chart conveys limited information. This is mainly reflected in the *determine range* task, where the chart only presents discrete variables. We think that all the reasons above are mainly aroused by the dataset, while we pay more attention to the form in which the data is presented. On the other hand, the abnormality of charts can also mirror the abnormality of the dataset, which is also meaningful in data analysis. So we believe that TaskVis still has a good recommendation ability, and we will also handle the problem in future work.

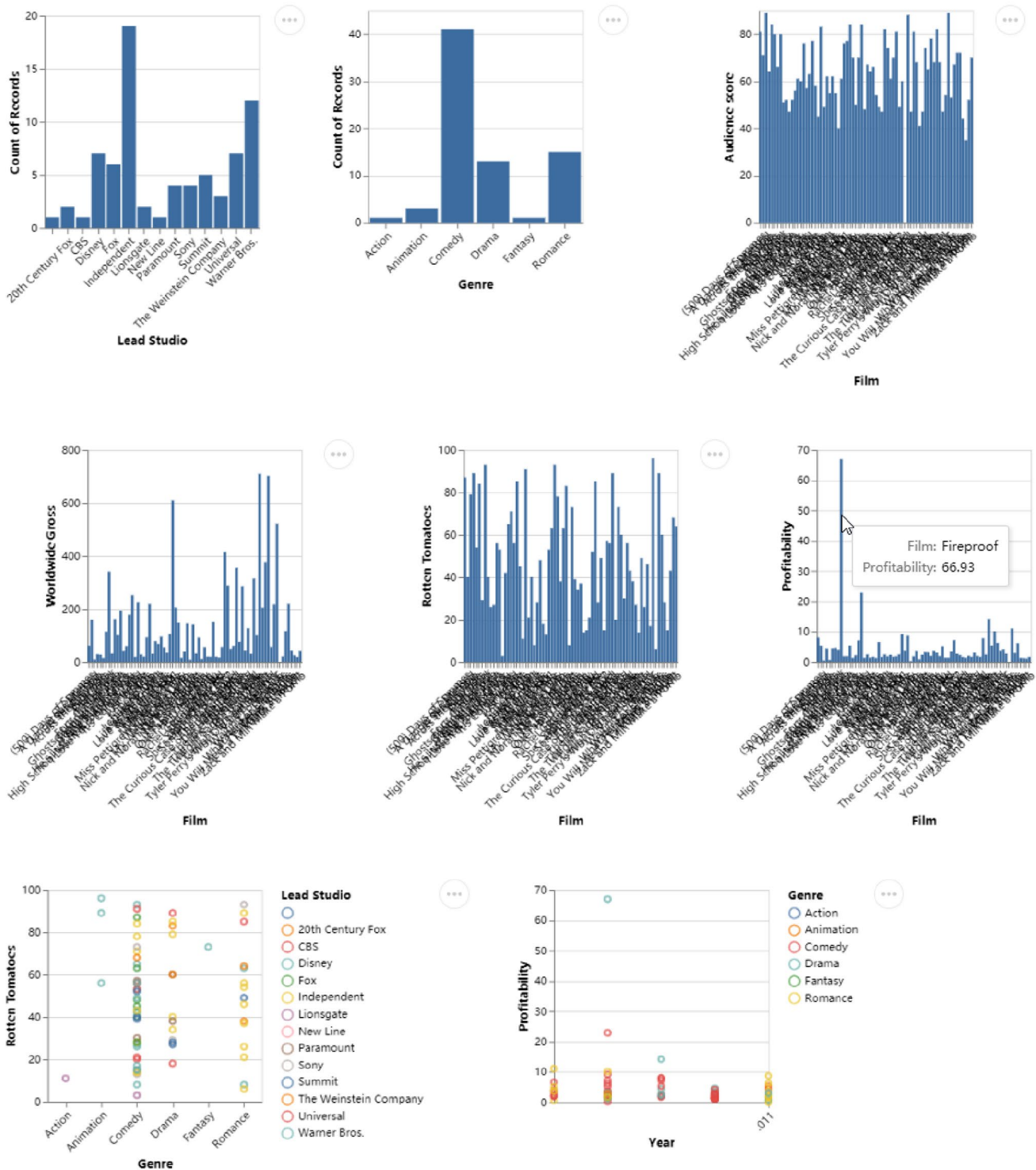

**Fig. 9** Combination Recommendation results of the Hollywood Stories dataset (without task selection)

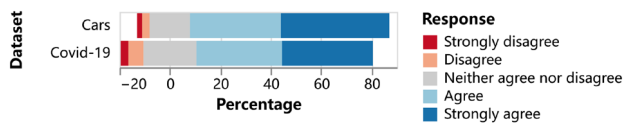

**Fig. 10** User ratings of whether the visualization can satisfy the given task with a 5-point Likert scale

In the post-study interviews, TaskVis was generally considered usable. Participants mostly agreed with the statements, "*I think that I would like to use this system frequently.*" ($\mu$ = 4.3 on a 5-point Likert scale) and "*I thought the system was easy to use*" ($\mu$ = 4.5 on a 5-point Likert scale). The mean system usability score is $\mu$ = 71, being between "OK" and "Good" [8]. We also collected many feedbacks from users. P4 (female, 27) said that "*the integrated tasks cover commonly used tasks in my daily work and it (TaskVis) can help me quickly generate appropriate visualizations*". For the ranking effectiveness, most participants agreed that "*the ranking schemes can help me explore the recommended charts in a proper order*" ($\mu$ = 4.2 on a 5-point Likert scale). P1 (male, 23) made a comment that "*the four ranking schemes are interesting. I can choose different schemes according to my application scenarios. Usually, the charts listed in the front are relatively more appropriate*". P18 (female, 23) commented regarding the final presentation of the system that "*multiple presentation modes allow flexible exploration.*" When asked about the interaction design, P12 (male, 26) commented that "*the interface is easy to use. I can input data columns and tasks freely, and the presentation of generated charts is intuitive*". In addition, most users agreed that "*the combination recommendation strategy can help give a brief view of the data*" ($\mu$ = 4.2 on a 5-point Likert scale). P9 (male, 23) said that "*combination recommendation is useful, sometimes I select multiple tasks, but individual visualization may fail to cover all the tasks*". Some participants also pointed out the drawbacks of TaskVis. P13(male, 25) said that "*it would be better to make the output editable.*" and P14 (male, 23) stated that "*more data insights can be further integrated. More convenient interaction paradigms may enhance the overall user experience, such as pen, touch, and natural language.*" Overall, participants all agreed that TaskVis is an effective visualization tool to facilitate data exploration.

### 4.3 Comparison

TaskVis is a task-oriented VisRec system. There is no ideal technology in the community to compare with. So this subsection only provides qualitative discussions. To our knowledge, most existing VisRec systems can not accept the user's analytic tasks and fail to recommend targeted results. Instead, TaskVis pays more attention to the user's tasks and can generate richer charts. So the recommended visualizations of TaskVis can better satisfy the user's diverse analysis intents. Besides, most VisRec systems generate only a few charts in a recommendation circle. Instead, TaskVis can recommend various visualizations for users to select, and the charts ranked in the front are relatively more appropriate. Different ranking schemes also apply to different scenarios.

Recently, Visualization-oriented Natural Language Interfaces (V-NLIs) are becoming a complementary input modality to traditional WIMP interaction [50]. V-NLIs can extract the user's task by understanding the NL utterances since they may hint at the user's analysis goals. However, most V-NLIs can only deal with well-structured queries. They still face various challenges, such as the ambiguous and underspecified nature of human language. Compared with these technologies, TaskVis provides a set of commonly used analytic tasks and design an interaction to allow users to select their tasks precisely. Besides, the task base can be extended with more tasks in the future, and the system can easily integrate new tasks. Finally, we argue that TaskVis is not always superior to the aforementioned technologies. Rather, it provides a new approach to integrate the user's analytic tasks into VisRec systems and builds a task-oriented system that can generate targeted visualizations. We believe that these technologies and TaskVis can complement each other.

## 5 Related Works

TaskVis draws upon prior efforts in analytic task construction, visual design knowledge, and visualization recommendation systems.

### 5.1 Analytic Tasks

There is a growing body of literature recognizing that the analytic task is vital for visualization design. Vartak *et al.* [57] proposed that the intended task is one of the five factors that must be accounted for while making visualization recommendations. There have been numerous task taxonomies in the data visualization literature providing definitions of analytic tasks. For instance, Roth *et al.* [45] considered integrating the user's domain-independent goals in viewing data with graphical designs first. Amar *et al.* [13] presented a set of ten low-level analytic tasks that capture the user's activities while employing visualization tools for data exploration, including *retrieve value*, *filter*, *compute derived value*, *find extremum*, *sort*, *determine range*, *characterize distribution*, *find anomalies*, *cluster*, and *correlate*. Saket et al. [46] evaluated the effectiveness of five commonly used chart types (mark as table, line, bar, scatter, and pie) across the ten tasks [13] by a crowdsourced experiment. Kim et al. [31] measured subject performance across task types derived from the ten tasks mentioned above and included *compare values* task in addition. The tasks were further grouped into two categories: value tasks that just retrieve or compare individual values and summary tasks that require identification or comparison of data aggregation. Draco [39] and Dziban [33] are the following VisRec systems that involve value tasks and summary tasks. NL4DV [40] additionally included a *Trend* task. AutoBrief [30] further enhanced VisRec systems by introducing domain-level tasks. Wang et al. [60] summarized 11 categories of tasks that are commonly adopted in fact sheets. Deep into scatterplots, Sarikaya et al. [47] first collected model tasks from a variety of sources to formulate the seeds for a scatterplot-specific analytic task list, then performed card sort to group tasks together based on their similarity. Rather than inputting tasks proactively, behavior-based systems represented by HARVEST [23] and Eye Gaze [55] capture users' interaction behavior to infer their analytic task. Besides, various empirical materials aiming at improving chart literacy have gradually formed in the industry practice. For instance, choosing a good chart [4] identified four tasks for market researchers and Financial Times Visual Vocabulary [11] summarized nine tasks commonly used by designers and journalists. However, in the data analysis community, the design knowledge is rarely used to power VisRec systems.

### 5.2 Design Knowledge and Visualization Recommendation

Early VisRec works are mostly rule-based (knowledge-based). For instance, APT [37] maintained about 200 visual design rules in a logic language form to enumerate the encoding space. SAGE [44] extended APT by taking data features (e.g., cardinality and uniqueness) into account to prune large search space. Rank-by-Feature Framework [49] leveraged selected metrics to rank histograms and scatterplots. Show me [38] proposed a set of heuristic rules to suggest the construction of small multiples with appropriate mark types. The design rules in the community increase iteratively as subsequent works embody the previous works and produce new rules. Voyager [61, 62], DIVE [27], Polaris [56], Profiler [29] all contributed to the enrichment of rules with more data types. Instead of considering common visualization types, Wang et al. [59] developed rules for the automatic selection of line graphs or scatterplots to visualize trends in time series, while Sarikaya et al. [47] focused on scatter charts. Datesite [17] proactively generated insights using automatic algorithms for data exploration. CompassQL [61] offers a general-purpose partial specification language that describes enumeration constraints for choosing, ranking, and grouping charts. To better benefit from design guidance provided by empirical studies, Moritz et al. [39] proposed a formal framework that models visualization design knowledge as a collection of ASP (Answer Set Programming) constraints [22]. We also merge this idea into TaskVis and make further improvements.

Rule-based approaches highly depend on domain expertise. In response, learning-based systems draw from existing well-designed visualizations to train neural models. DeepEye [34] trained a decision-tree-based binary classifier to determine whether a visualization is good or not, and a learning-to-rank model to rank visualizations. Data2Vis [20] built a multi-layered attention-based encoder-decoder network to directly map JSON-encoded datasets to Vega-Lite specification. Draco-Learn trained a RankSVM model on ranked pairs of visualizations and then automatically learned weights for soft constraints in ASP. VizML [25] identified five key design choices while creating visualizations. For a new dataset, the 841 dataset-level features extracted were fed into neural network models to predict the design choices. Table2Charts [64] used a deep Q-learning approach with the copying mechanism and heuristic searching strategy. Qian et al. [41] proposed an end-to-end machine learning approach while MultiVision [63] built a Siamese neural network. However, these learning-based approaches can not achieve great performance with the limited training data, and the large datasets are usually unavailable in the community. So this paper focuses on rule-based methods

## 6 Conclusion and Future Work

In this paper, we present a task-oriented visualization recommendation approach called TaskVis. TaskVis improves the existing technologies by strongly correlating recommendations with the user's analytic task. It allows users to input their tasks and interests flexibly, thereby striking a better balance between automation and the user's intent. We hope that it can inspire more valuable works in the VisRec community. Nevertheless, it is still in the early stage of visualization recommendation. We now discuss how TaskVis can evolve to be more intelligent and what we need to do in the future.

*Extract abundant data properties* The rules in TaskVis are more detailed to the task but less involved in data properties, where only the type, length, and cardinality of data columns are considered. However, the properties of multi-dimensional statistical data are beneficial for targeted recommendations, such as standard deviation, coefficient of variance, and quantitative coefficient of dispersion extracted in VizML [25]. In the future, we will extend the rule base to support more statistical data properties.

*Integrate semantics and natural language interface* TaskVis only integrates data properties in the numerical sense but does not consider the semantic meaning. In the future, we will integrate the semantic type of data columns into TaskVis to transmit more valuable information [28]. Further, the user's occupation, social hotspots, and geographic information can be helpful to the personalized recommendation. In addition, a natural language interface can also greatly improve the user experience of the VisRec systems [35, 36, 50].

*Knowledge graph to construct rules* Draco [39] has formalized visualization design knowledge as constraints, but the knowledge is independent of each other, which is not conducive to knowledge reasoning. We suppose leveraging a knowledge graph to construct a rule base in an extensible model where the rules are interrelated. Based on the Knowledge graph, we can conduct knowledge extraction from public examples. In addition, users can also be allowed for customizing rules to satisfy their requirements.